\documentclass[prd,aps,amsmath,twocolumn,superscriptaddress,amssymb,reprint,nofootinbib]{revtex4-2}

\usepackage{graphicx,array}
\usepackage{hyperref}
\usepackage{color}
\usepackage{amsmath,amssymb,slashed,latexsym}
\usepackage{bm}
\usepackage{braket}
\usepackage{placeins}
\usepackage{subcaption}
\usepackage{enumitem}

\def\beq{\begin{equation}}
\def\eeq{\end{equation}}
\def\bea{\begin{eqnarray}}
\def\eea{\end{eqnarray}}

\definecolor{darkblue}{cmyk}{1,0.4,0,0.3}
\definecolor{violet}{cmyk}{0,1,0,0.2}
\hypersetup{colorlinks, bookmarksnumbered, citecolor=darkblue, linkcolor=darkblue, pdfstartview=FitH, urlcolor=darkblue, linktocpage}

\begin{document}

\title{Attracting the Electroweak Scale to a Tachyonic Trap} 

\author{\large Sokratis Trifinopoulos}\email{trifinos@mit.edu}
\affiliation{Center  for  Theoretical  Physics,  Massachusetts  Institute  of  Technology,  Cambridge,  MA  02139,  USA}

\author{\large Miguel Vanvlasselaer}\email{miguel.vanvlasselaer@vub.be}
\affiliation{Theoretische Natuurkunde and IIHE/ELEM, Vrije Universiteit Brussel, \& The  International Solvay Institutes, Pleinlaan 2, B-1050 Brussels, Belgium}
\affiliation{SISSA International School for Advanced Studies, Via Bonomea 265, 34136, Trieste, Italy}
\affiliation{INFN, Sezione di Trieste, SISSA, Via Bonomea 265, 34136, Trieste, Italy}

\begin{abstract}
We propose a new mechanism to dynamically select the electroweak scale during inflation. An axion-like field $\phi$ that couples quadratically to the Higgs with a large initial velocity towards a critical point $\phi_c$ where the Higgs becomes massless. When $\phi$ crosses this point, it enters a region where the Higgs mass is tachyonic and this results into an explosive production of Higgs particles. Consequently, a back-reaction potential is generated and the field $\phi$ is attracted back to $\phi_c$. After a series of oscillations around this point it is eventually trapped in its vicinity due to the periodic term of the potential. The model avoids transplanckian field excursions, requires very few e-folds of inflation and it is compatible with inflation scales up to $10^5~\rm GeV$. The mass of $\phi$ lies in the range of hundreds of GeV to a few TeV and it can be potentially probed in future colliders.

\end{abstract}

\maketitle

\section{Introduction}
\label{sec:intro}

In recent years, the idea that the electroweak scale could be dynamically determined by the cosmological evolution of a (pseudo-)scalar field sparked a paradigm shift in theories of naturalness. The first model of this kind~\cite{Graham:2015cka} features an axion-like field $\phi$, called the relaxion, which couples to the Higgs $H$ via a term of the type $g \Lambda \phi H^2$ with tiny $g$. The relaxion slow-rolls during inflation and scans the Higgs mass $m_H^2(\phi)=-\Lambda^2+g \Lambda \phi$, where $\Lambda$ is the scale that New Physics (NP) is expected to appear. Electroweak symmetry breaking occurs after the field crosses the critical point $\phi_c = \Lambda /g$ and a periodic back-reaction potential for $\phi$ is generated via non-perturbative effects of a confining sector at scale $M$. The height of the potential barriers grows with the increasing Higgs vacuum expectation value (VEV), eventually stopping the relaxion and trapping it into a local minimum at the electroweak scale $v_{\rm EW}$. No new degrees of freedom at the TeV scale charged under the Standard Model (SM) are required and as a result experimental strategies motivated by naturalness are radically different in this framework. 

The original proposal was not without some theoretical shortcomings such as the requirement $M \lesssim v_{\rm EW}$, which implies that the confining sector is hidden (i.e. not charged under the SM symmetries) and its scale coincides with the electroweak scale without any a priori reason. Moreover, transplanckian field excursions of the relaxion $\Delta \phi \sim \Lambda /g$ are necessary as well as an enormous number of e-folds that have to be produced by low-scale inflation, which raises concerns of cosmological fine-tuning~\cite{Hardy:2015laa,Fowlie:2016jlx}. Various model-building attempts to address these issues have appeared in the literature~\cite{Espinosa:2015eda,Hardy:2015laa,Gupta:2015uea,Choi:2015fiu,Kaplan:2015fuy,Ibanez:2015fcv,Hebecker:2015zss,Fonseca:2016eoo,Evans:2016htp,Hook:2016mqo,Choi:2016kke,Tangarife:2017rgl,Wang:2018ddr,Fonseca:2018xzp,Ibe:2019udh,Fonseca:2019lmc,Domcke:2021yuz,Klangburam:2022igc,Chatrchyan:2022pcb}, albeit at the price of introducing non-minimal setups. Beyond the relaxion framework, recent works~\cite{Geller:2018xvz,Cheung:2018xnu,TitoDAgnolo:2021nhd} have considered scenarios in which the electroweak scale is also determined due to the interplay between a scalar and the Higgs, but instead of a dynamical relaxation there is environmental and anthropical selection related to the vacuum energy in different patches of the inflationary universe. 

In this letter we present a model of cosmological relaxation of the electroweak scale which is free of the above-mentioned pathologies while at the same time remains economical introducing only one new field at the effective theory level. In particular, it utilizes a stopping mechanism that relies on the extremely rapid production of excitations of a scalar field, in our case the Higgs field, that couples quadratically to another (pseudo-)scalar field $\phi$. The particle production takes place when the Higgs becomes massless at a critical point of the classical trajectory of $\phi$, i.e. the symmetry breaking point (SBP) $\phi_c$. The produced particles generate an effective back-reaction potential that attracts the field $\phi$, which we will call \emph{attraxion}, back to the SBP. If the production is strong enough, the global minimum of the potential is now $\phi = \phi_c$ and the field starts to oscillate around it. Hubble expansion causes a decrease of the oscillation amplitude and eventually the field is trapped in the vicinity of the SBP. A similar trapping mechanism was first envisioned as a possible solution to the cosmological moduli problem~\cite{Kofman:2004yc} and then exploited in models of \emph{trapped inflation}~\cite{PhysRevD.80.063533,Pearce:2016qtn} as a method to obtain slow-rolling conditions for the inflaton even in a non-flat potential. More recently it has been used in the context of quintessential inflation in order to freeze the inflaton dynamics until later times~\cite{Dimopoulos:2019ogl,Karciauskas:2021fdu}.

In contrast to the slow-rolling relaxion, the mechanism is effective in the high initial velocity regime of the parameter space, which additionally enables a fast scanning of the Higgs mass requiring only very few e-folds of inflation. The attraxion potential also has a periodic term which is initially not interfering with the fast rolling, but after the kinetic energy is depleted, the field is eventually trapped in one of its valleys. The process occurs before the Higgs number density is diluted due to inflation or the Higgs bosons decay removing the back-reaction term. It is worth noticing that the periodic potential does not depend on the Higgs VEV disentangling in principle the scale of the confining sector from the electroweak scale. Furthermore, the size of the coupling $g$ required by the mechanism is much larger than the one in relaxion models which implies that field excursions are always smaller than the Planck scale.

\section{The attraxion model}
\label{sec:model}

\textbf{\emph{Effective potential}} - The effective potential at tree-level reads
\bea \label{eq:pot_tree}
 V_{\rm tree}(H,\phi) = g^2 \frac{\phi^2 -\phi_c^2}{2} |H|^2
 +\frac{\lambda}{4}|H|^4 +V_\phi(\phi)~,
\eea 
where $\phi_c \equiv \Lambda/g$. The potential has two SBPs at $\phi = \pm \phi_c$. In this letter, we study the case of a quadratic attraxion-Higgs coupling (e.g. see Ref.~\cite{Gupta:2015uea}).

We assume that $\phi$ does not couple at tree level to the NP at scale $\Lambda$. Despite that, closing the Higgs loop provides the leading loop-level correction
\bea \label{eq:pot_loop}
V_{\rm loop}(\phi) \sim \frac{g^4 \phi_c^2}{16\pi^2} \phi^2 = \frac{g^2 \Lambda^2}{16\pi^2} \phi^2~.
\eea 

Finally, we assume that $\phi$ obeys a shift symmetry broken at scale $f$ and couples to a hidden confining sector at scale $M$, which yields the periodic potential
\bea \label{eq:pot_periodic}
V_{\phi}(\phi) =  M^4 \cos \frac{\phi}{ f}~.
\eea 
Unlike in the traditional relaxion model, this term does not depend on the Higgs VEV and is present even before the stopping mechanism is triggered. This term allows for the existence of local minima close to the SBPs when 
\bea \label{eq:min_existence}
M^4 \gtrsim \frac{g \Lambda^3 f}{8\pi^2}~.
\eea
A concrete ultraviolet (UV) completion is beyond the scope of this letter, but we mention that variations of the constructions laid out in Refs.~\cite{Graham:2015cka,Antipin:2015jia,Gupta:2015uea} and in particular the clockwork framework of Ref.~\cite{Kaplan:2015fuy,Choi:2015fiu} could match to our model in the low-energy limit.

The electroweak symmetry is broken in the region $\phi<\phi_c$, where the minimum of the potential in the Higgs direction is situated at 
\bea \label{eq:higgs_minimum}
v_H^2(\phi) = \frac{g^2}{\lambda}(\phi_c^2-\phi^2)~.
\eea 
The minimum of the potential in the attraxion direction is at $\phi=0$.

\textbf{\emph{Trapping mechanism}} - The rolling of the attraxion starts during the inflation era at large negative field values $\phi_i \ll -\phi_c$ (the choice of the sign is free) and with a large initial velocity towards the origin. As the attraxion comes close to the first SBP $\phi=-\phi_c$ with velocity $\dot \phi_c$, the Higgs becomes massless. The Higgs modes with momentum $k$ and frequency $\omega_k = \sqrt{k^2+m_H^2(\phi)}$ for which the non-adiabatic parameter $\dot \omega_k/\omega_k^2$ becomes large, are excited and resonant particle production takes place~\cite{Kofman:1997yn}. After it crosses the SBP, the mass parameter becomes negative and the modes with $k^2<|m_H^2(\phi)|$ will be exponentially amplified via a process called \emph{tachyonic resonance}~\cite{Dufaux:2006ee,Dufaux:2008dn,Abolhasani:2009nb,Fedderke:2014ura,Karciauskas:2021fdu}. The particle production occurs throughout the non-adiabatic region between the two SBPs $|\phi|<\phi_c$ and it peaks at $\phi=0$, where the maximal number of modes become tachyonic.

The Higgs quartic self-interaction $\lambda h^4$ reintroduces an effective mass term $m_H^2+3\lambda \langle H^2 \rangle_{\rm eff}$, which suppresses the particle production. Taking this effect into consideration, in Ref.~\cite{Karciauskas:2021fdu} the authors derive an analytic approximation for the total particle number density after the exit from the non-adiabatic region at the second SBP $\phi=\phi_c$,
    \bea     \label{eq:higgs_number}
    n_{H} \approx \left(\frac{\sqrt{g\dot \phi_c}}{2\pi}\right)^3 e^{\frac{\pi \Lambda^2}{g\dot \phi_c}}
    \times e^{-\frac{3\pi\lambda(\Lambda){\langle H^2\rangle}_{\rm eff}^{(0)}}{g\dot \phi_c}} ~,
    \eea 
    where
    \bea     
    {\langle H^2\rangle}_{\rm eff}^{(0)}=\frac{g \dot \phi_c}{2\pi^3} \sqrt{\frac{\pi/2}{\big|1-Q/2 \big|}} e^{Q/2-1}~, \ Q\equiv \frac{\pi \Lambda^2}{g \dot \phi_c}~.
    \eea 
The production is favored for smaller values of the quartic. Notice that $\lambda$ is evaluated at scale $\Lambda$, because this is the relevant energy scale of the Higgs potential at the point of maximum production. In the following and unless explicitly mentioned otherwise, we will abbreviate $\lambda=\lambda(\Lambda)$ and consider it as a free parameter.

The corresponding energy density stored in Higgs excitations is~\cite{Kofman:2004yc}
    \bea \label{eq:rho_H}
    \rho_H &\approx n_H |m_H(\phi)| \approx \begin{cases}
 \sqrt{2}g n_H |\Delta \phi|~, \quad |\Delta \phi| \gg \phi_c
\\
 g n_H \sqrt{2 \phi_c |\Delta \phi|}~, \quad |\Delta \phi| \ll \phi_c
\end{cases}~, \notag \\
    \eea 
   where $\Delta \phi = \phi-\phi_c$. 
   
   For a wide range of model parameters we have
   \bea 
  V(0) > V(\phi_c) ~ \Rightarrow ~
  n_H |m_H(0)| - \frac{\Lambda^4}{4\lambda} > \frac{\Lambda^4}{16\pi^2}~,
  \eea   
which implies that $\rho_H$ acts as a back-reaction potential and the SBP $\phi=\phi_c$ becomes the new global minimum attracting $\phi$ back to it. In fact, as the attraxion moves away from the SBP its kinetic energy is transferred to the Higgs energy density and when $\rho_H \sim \dot \phi_c^2/2$ at $\Delta \phi=A$ with
    \bea \label{eq:initial_amp}
    A \equiv \begin{cases}
 \frac{\sqrt{2} \dot \phi_c^2}{4 g n_H}~, \quad |\Delta \phi| \gg \phi_c
\\
 \frac{\dot \phi_c^4}{8 g n_H^2 \Lambda}~, \quad |\Delta \phi| \ll \phi_c
\end{cases}~,
    \eea
it stops and returns back to the SBP. As it crosses this point again (with practically the same velocity) it triggers a second burst of particle production and the newly created Higgs bosons are added to the total bath.\footnote{In the region of the parameter space that is interesting for our setup, we know a posteriori that the maximum particle production is achieved by the first burst. Since this induces a large term $\langle H^2 \rangle_{\rm eff}$, it either stops the particle production immediately after the first oscillation or renders the rest of them subdominant.} The attraxion dynamics enter a phase characterized by fast oscillations around the SBP.\footnote{Note that this phase is unique in our setup. Models that follow the slow-rolling relaxion paradigm and utilize particle production triggered at the SBP (e.g. see Ref.~\cite{Hook:2016mqo}) use the effect as a friction term, while the relevant term in our case corresponds to a restoring  force towards the SBP.} The Hubble friction dilutes the Higgs number density and dissipates the kinetic energy. As a consequence the amplitude of each oscillation $A(t)$ and the velocity $\dot \phi_c(t)$ at the SBP\footnote{In the text, unless otherwise explicitly mentioned, we denote as $\dot \phi_c$ the velocity during the first passage from the SBP.} both decrease with time (see Appendix \ref{app:redshift}).

Eventually, the kinetic energy of the attraxion drops enough so that the periodic potential (see Eq.~\eqref{eq:pot_periodic}) becomes relevant. The oscillations will stop in the local minimum closest to the SBP $\phi_{\rm min} \sim \phi_c -  f$. The Higgs field which was initially anchored at the origin $\langle H \rangle = 0$ now acquires the VEV
    \bea \label{eq:EW_VEV}
    v_H^2(\phi_{\rm min}) = v_{\rm EW}^2~.
    \eea 
The attraxion has to remain there until the present time. 
The mass of the attraxion is given by
\bea 
 m_{\phi}^2 = \frac{\partial^2V}{\partial\phi^2}\bigg|_{\phi=\phi_{\rm min} } \approx \frac{M^4}{f^2}+ \frac{g^2 \Lambda^2}{8\pi^2} - \frac{2 g^2 \Lambda^2}{\lambda(v_{EW})}~.
\eea    
Unlike in the relaxion case, as we will see, this is typically larger than the mass of the Higgs. In this limit, we may also write the mixing angle between the two scalars as
\bea 
\sin\theta \approx \left({\frac{\partial^2V}{\partial H\partial\phi}\Big/\frac{\partial^2V}{\partial\phi^2}}\right)\bigg|_{\langle H \rangle =v_H}^{\phi=\phi_{\rm min}} \approx \frac{2g^2}{m_{\phi}^2}\sqrt{\frac{2 \Lambda^3 f}{\lambda(v_{EW})}}~. \quad
\eea 

\section{Conditions for successful trapping}
\label{sec:conditions}

In this Section, we investigate further the details of the trapping mechanism by listing the necessary conditions for its realization in ``chronological order''.

\begin{enumerate}
\item {{\bf Classical over quantum.}}    
    The attraxion evolution must be dominated by classical rolling and not by the quantum fluctuations during inflation:
    \bea \label{eq:cvsq}
    \dot \phi_c > H_{\rm inf}^2~.
    \eea
   
\item {{\bf Inflaton domination.}}    
    For inflation to occur the energy budget must be dominated by the inflaton potential energy and not by the kinetic energy of the attraxion:
    \bea \label{eq:inflaton_dom}
    \dot \phi_c^2 \ll H_{\rm inf}^2M^2_{\rm pl}~.
    \eea

\item { {\bf Selecting the electroweak scale.}}
    The final trapping should occur in a valley of the periodic potential close to the SBP (see Eq. \eqref{eq:EW_VEV}). In order to satisfy Eq. \eqref{eq:EW_VEV} we require then   
    \bea \label{eq:EW_selection}
    f\sim \frac{\lambda(v_{EW}) v_{\rm EW}^2}{g\Lambda}~.
    \eea

\item {{\bf Efficient trapping.} } The trapping is achieved at the right scale if at time $t_{\rm EW}$ the amplitude of the oscillation enters the region
\bea 
\label{eq:condition_trap}
A(t_{\rm EW}) \lesssim 2 \pi f~.
\eea
Afterwards, the Higgs number density is diluted due to inflation to the point that the periodic term in the potential takes over and the minima, previously erased by the back-reaction term, reemerge. This condition is equivalent to equating the slopes of the two terms, i.e.
\bea \label{eq:trapping_ins}
\frac{M^4}{f} \gtrsim g n_H(t_{\rm trap}) \sqrt{\frac{\phi_c}{2 A(t_{\rm trap})}}~.
\eea
As shown in Appendix \ref{app:redshift} the amplitude close to the SBP ``red-shifts'' as $A \propto a^{-6/5}$. 
If $t_1$ is the time when the attraxion reaches the amplitude of the first oscillation, we can calculate the number of necessary e-folds as
\bea 
\qquad \frac{a(t_{\rm trap})}{a(t_1)} = \left(\frac{A(t_1)}{A(t_{\rm trap})} \right)^{5/6} =
\left( \frac{\dot \phi_c^4}{8 g n_H^2 \Lambda A(t_{\rm trap})}\right)^{5/6}~. \quad
\eea
The produced abundance of Higgs excitations redshifts like matter $n_H \propto a^{-3}$. Solving now Eq. \eqref{eq:trapping_ins} for $A(t_{\rm trap})$ and requiring that $a(t_{\rm trap})>a(t_{\rm EW})$ yields the following bound for the particle production
\bea \label{eq:trapping_amp}
n_H(t_1) > \frac{\sqrt{2}M^{2/3}\dot \phi_c^{5/3}}{4 \pi^{1/3}  f^{1/2}g^{1/2} \Lambda^{1/2}}~.
\eea 
An effect that can disturb the trapping is also the perturbative Higgs decay, which suppresses $n_H$ further. The decay can be neglected if its rate is slower than the Hubble expansion rate $\Gamma_H < H_{\rm inf}$. The maximum Higgs decay rate
\bea \label{eq:higgs_decay}
\Gamma_H^{\rm max} \approx \frac{y^2_t |m_H(A)|}{16\pi} = \frac{y^2_t}{32\pi }\frac{\dot \phi_c^2}{ n_{H} }~,
\eea 
provides then a lower bound on the inflation scale.

\item { {\bf No Freezing before trapping.}}
In the derivation of the evolution of the oscillation amplitude in the Appendix \ref{app:redshift}, we assume that the Hubble expansion is negligible during the timescale of one oscillation. However, this approximation breaks down at time $t_{\rm freeze}$ when the Hubble friction term is comparable with the slope of the back-reaction potential and the dynamics freeze
\bea \label{eq:higgs_decay}
 3H_{\rm inf} \dot \phi_{c,\rm f} \sim g n_H(t_{\rm freeze}) \sqrt{\frac{\phi_c}{2 A(t_{\rm freeze})}}~,
\eea 
where $\dot \phi_{c,\rm f}$ is the velocity at the last passage via the SBP before the time $t_{\rm freeze}$.
By requiring that $A(t_{\rm freeze}) \lesssim 2\pi f$ and repeating the same steps that lead to Eq. \eqref{eq:trapping_amp} we find the following upper limit for the inflation scale
\bea \label{eq:freezing}
H_{\rm inf} < \frac{\dot \phi_c}{24 \pi f }~.
\eea 

\item { {\bf Stability of the minimum during inflation}}:
During inflation, the space-time has a de Sitter geometry, which is known to mimic thermal effects with fluctuations of order $\frac{H_{\rm inf}}{2\pi}$. Those effects would destabilise the trapping minimum unless 
   \bea  \label{eq:inflation_fluc}
   H_{\rm inf} < 4\pi^2 f~.
   \eea 
\end{enumerate}

Additionally, we mention that the trapped minimum represents a metastable vacuum, which could undergo quantum tunnelling (see Appendix \ref{sec:tunneling} for the calculation of the transition rate). However, we find that vacuum stability until today is ensured in all the relevant part of the parameter space.

\section{Parameter space and future prospects}
\label{sec:results}

\begin{figure}
\centering
\includegraphics[scale=0.6]{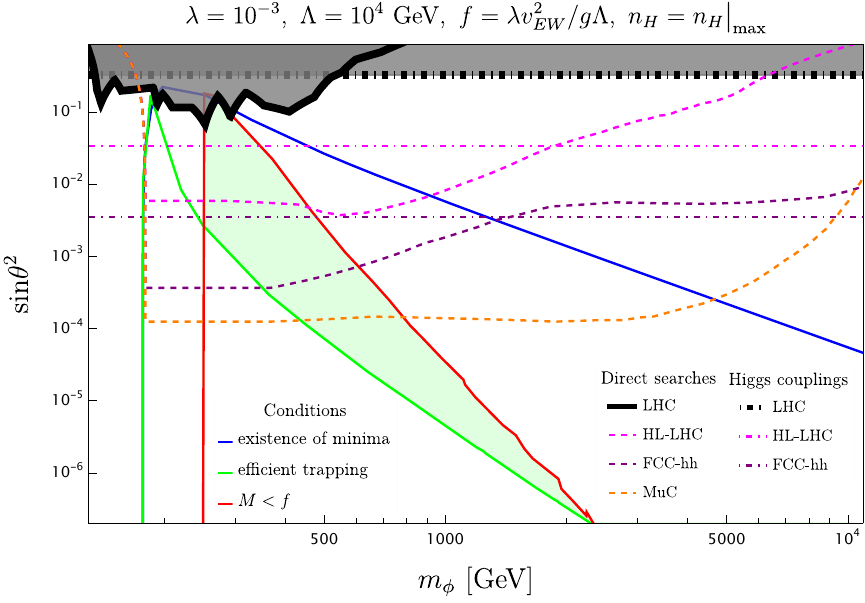}
\caption{Parameter space of the model. In the green region all the cosmological conditions are satisfied. Current exclusion limits at LHC as well as direct and indirect reach at future colliders are shown at 95\% CL.}
\label{fig:parameter_space}
\end{figure}

\textbf{\emph{Charting the viable parameter space}} - The model parameters are the couplings $g$, $\lambda$, the scales $f$, $\Lambda$, $M$ and the initial velocity $\dot \phi_c$. In the analysis we fix $f$ according to Eq. \eqref{eq:EW_selection} and express $\dot \phi_c$ as a function of $g$ by requiring that the Higgs particle number density $n_H$ given in Eq. \eqref{eq:higgs_number} is maximized. We find that this happens for velocity values around  $\dot \phi_c \approx c \Lambda^2/(10 g)$, where $c$ is an $\mathcal{O}(1)$ parameter that depends on the choice of $\lambda$. For the Higgs quartic and the the NP scale we use the benchmark $\lambda(\Lambda)=10^{-3}$ and $\Lambda=10^4~ \rm GeV$, respectively.

The parameter space that realizes the trapping mechanism can then be presented in a two-dimensional plane. We employ the parameters $\sin\theta^2$ and $m_\phi$ and our results can be found in Fig. \ref{fig:parameter_space}. The most stringent bounds are imposed by Eq. \eqref{eq:min_existence} (blue), Eq. \eqref{eq:trapping_amp} (green) and the expectation $M < f$ (red) from the axion-like effective theory construction.

\textbf{\emph{Collider bounds and prospects}} - The attraxion couples to the SM particles via its mixing with the Higgs which is proportional to $\sin\theta$. An upper bound of $\sin\theta \lesssim 0.37$ (dash-dotted black) is obtained by indirect measurements of the SM-like Higgs couplings~\cite{ATLAS:2016neq,Ilnicka:2018def,Adhikari:2020vqo}. HL-LHC (dash-dotted  magenta) and FCC-hh (dash-dotted  purple) are expected to improve on this bound~\cite{EuropeanStrategyforParticlePhysicsPreparatoryGroup:2019qin} by one and two orders of magnitude, respectively. 

Direct searches are also relevant, since the attraxion can be singly produced via vector boson fusion and then decay to a pair of SM gauge bosons or SM-like Higgs bosons $\phi \to ZZ$ (or $hh$). Present LHC exclusion limits~\cite{ATLAS:2017otj,CMS:2018amk} (dashed black) are not constraining, while HL-LHC (dashed magenta) will be able to probe masses up to $500 ~\rm GeV$ for $\sin\theta \gtrsim 0.07$. Regarding future colliders, a 14 TeV MuC (dashed orange) offers better sensitivity than FCC-hh (dashed purple) constraining all relevant masses for $\sin\theta \gtrsim 0.01$~\cite{Buttazzo:2018qqp,EuropeanStrategyforParticlePhysicsPreparatoryGroup:2019qin,MuonCollider:2022xlm}. 

One can then directly map the bounds derived for generic Higgs-singlet portal models on the $\sin\theta-m_\phi$ plane. In Fig. \ref{fig:parameter_space}, we provide the current exclusion limits from LHC as well as the projections for the reach of HL-LHC, a $100~\rm TeV$ FCC-hh and a 14 TeV Muon Collider (MuC) at the 95\% CL. We infer that in the scenario where the NP cut-off lies at the $10~\rm TeV$ direct detection of the attraxion will be possible for a considerable part of the parameter space.

\begin{figure}
\centering
\includegraphics[scale=0.5]{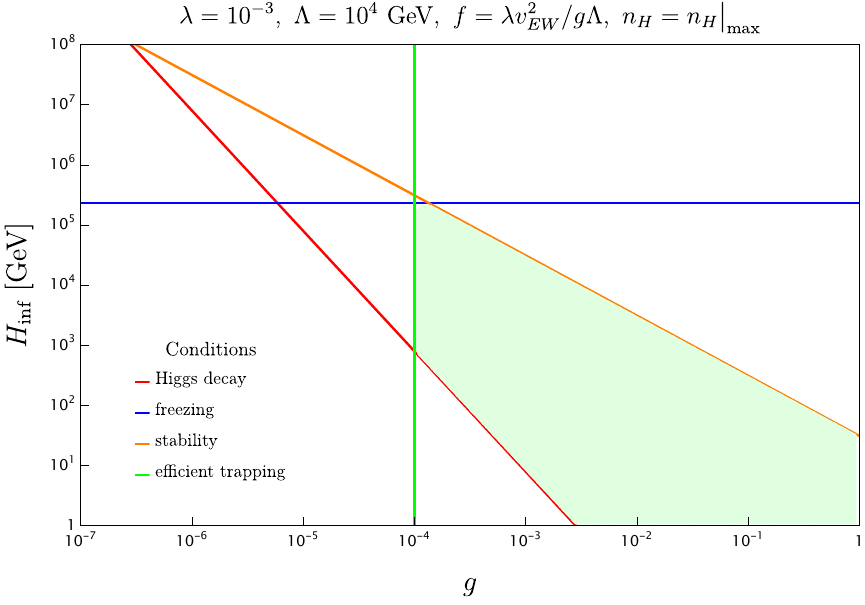}
\caption{Inflation scale compatible with the trapping mechanism.}
\label{fig:Hrange}
\end{figure}

\textbf{\emph{Inflation and New Physics scales}} - In Fig. \ref{fig:Hrange} the inflation scale is displayed as a function of the coupling $g$. The viable range is constrained by Eqs. \eqref{eq:higgs_decay} (red), \eqref{eq:inflation_fluc} (orange) and \eqref{eq:freezing} (blue), while Eqs. \eqref{eq:cvsq} and \eqref{eq:inflaton_dom} are readily satisfied. The rest of the conditions (that determine the allowed region in Fig. \ref{fig:parameter_space})  also yield a lower bound for $g$ (green). We observe that our mechanism allows for scales significantly higher than the case of the relaxion, with a maximum of order $10^5~\rm GeV$. Moreover, the completion of the trapping occurs after a modest number of e-folds $\log[ A(t_1)/2\pi f] \sim \log[ 10^{-4} \Lambda^4 / (g^2 n_H v_{EW})] \lesssim \mathcal{O}(10)$. As a result, since the high-velocity regime of the attraxion is realized at the onset of inflation, we expect that the whole trapping will be finished before the end of the main inflationary era.  

Regarding the upper bound on the NP scale, for $\lambda(\Lambda)=10^{-4}$ we find it to be around $200~\rm TeV$. In principle, much higher NP scales can be reached on a basis of an UV-motivated argument for a smaller Higgs quartic at that scale.

\section{Discussion and Conclusions}

In this letter, we propose a novel explanation for the smallness of the electroweak scale based on the dynamical evolution of the attraxion field $\phi$ during the era of inflation. Our construction shares conceptual similarities and the minimality of the original relaxion model~ transposed in a regime of high initial velocity. Nevertheless, the resulting dynamics and phenomenology are entirely different. In particular, it addresses the four basic conditions for constructing models of cosmological relaxation as outlined in Ref. ~\cite{Graham:2015cka} in the following fashion:

\begin{enumerate}[label=\roman*)]
\item Higgs back-reaction is achieved thanks to the rapid production of Higgs particles via tachyonic resonance at special points on the trajectory of $\phi$.
\item Dissipation of kinetic energy is achieved thanks to the energy transfer in the back-reaction sector and the subsequent dilution of the produced Higgs number density due to the inflationary expansion.
\item Self-similarity is a consequence of the axion-like nature of the attraxion. However, the height of the barriers remains constant.
\item A long period of scanning field evolution is no longer necessary since the attraxion is fast-rolling and the whole process is completed in less than 10 e-folds of inflation.
\end{enumerate}

Among models of cosmological relaxation, our proposal uniquely features a rather sizeable pseudoscalar-Higgs coupling and a pseudoscalar mass heavier than the Higgs mass. The model is thus realized without the requirement of transplanckian field space. Ultimately, the most promising experimental avenue for the detection of the new state $\phi$ become again collider searches. For the case of New Physics at the $\mathcal{O}(10-100)$ TeV scale, the attraxion can be directly probed at future colliders with the $14~\rm TeV$ Muon Collider offering the best sensitivity.

\begin{acknowledgments}
We thank Paolo Creminelli and Keith Olive for their comments on the early stages of the project. We also thank Aleksandr Azatov, Takeshi Kobayashi, Lorenzo Ubaldi and Geraldine Servant for their fruitful feedback.
ST is supported by the Swiss National Science Foundation - project n. P500PT\_203156, and by the Center for Theoretical Physics, Massachusetts Institute of Technology. MV is supported by the ``Excellence of Science - EOS" - be.h project n.30820817, and by the the Strategic Research Program High-Energy Physics of the Vrije Universiteit Brussel.
\end{acknowledgments}

\appendix 

\section{Evolution of the amplitude of the attraxion oscillations}
\label{app:redshift}

The equation of motion for the attraxion during the Phase 2 of the trapping mechanism reads
\bea \label{eq:EoM}
\ddot \phi + 3H\dot\phi + \frac{\partial V}{\partial\phi} = 0 
\eea 
where $V \approx \rho_H$ (see main text Eq. (10)) and we have replaced $\Delta \phi \to \phi$ for convenience. The energy density of the attraxion is 
\bea \label{eq:rho}
\rho = \frac{ \dot \phi^2}{2} + V= \frac{ \dot \phi_c^2}{2} = V(A)~.
\eea 
Over the time-scale of one oscillation, we can neglect the expansion of the universe and consider roughly $\rho \approx \text{const}$. Following Ref. \cite{PhysRevD.28.1243} we calculate the the average kinetic and potential energy as
\bea \label{eq:average_K_V}
\langle K\rangle  = \frac{\gamma}{2} \rho~, \quad \langle V\rangle = \frac{2-\gamma}{2}\rho~, 
\eea
where
\bea 
\gamma \equiv 2\frac{\int_{0}^{A} d \phi (1-V(\phi)/\rho)^{1/2}}{\int_{0}^{A}d \phi (1-V(\phi)/\rho)^{-1/2}}=\begin{cases}
 2/3~, \quad |\phi| \gg \phi_c
\\
 2/5~, \quad |\phi| \ll \phi_c
\end{cases} ~.
\eea 
Now we may consider the effect of the Hubble expansion imposed on the time evolution of the averaged quantites. Starting from Eq. \eqref{eq:EoM} and replacing Eq. \eqref{eq:rho} (after taking the derivative) we get
\bea \label{eq:eom_rho_1}
\dot \rho - \dot V +3H\dot\phi^2+\frac{\partial V}{\partial\phi} \dot \phi &= 0 \notag \\
\dot \rho + 3H\dot\phi^2 - V\frac{\dot n_H}{n_H}&=0~,
\eea 
where we have used $\dot V = \frac{\partial V}{\partial\phi}\dot\phi+V\frac{\dot n_H}{n_H}$. Averaging the above expression over an oscillation and using Eq. \eqref{eq:average_K_V} it follows 
\bea \label{eq:rho_redshift}
\dot \rho + 3H\gamma \rho - \frac{2-\gamma}{2} \frac{\dot n_H}{n_H} \rho&=0 \notag \\
\dot \rho + 3\frac{\gamma+2}{2} H\rho&=0~,
\eea 
where we have used that $\dot n_H =-3H n_H$.
We thus obtain 
\bea \label{eq:rho_redshift}
 \rho \propto a^{-3(\gamma+2)/2}~.
\eea
Given the definition of the amplitude according to Eq. (12) (main text) we finally infer for the evolution of the amplitude of the oscillation
\bea 
A \propto \begin{cases}
 a^{-1}~, \quad |\phi| \gg \phi_c
\\
 a^{-6/5}~, \quad |\phi| \ll \phi_c
\end{cases}~.
\eea 

\hspace{1cm}

\section{Quantum tunnelling}
\label{sec:tunneling}

At late times the trapped minimum represents a metastable vaccuum, which could undergo quantum tunneling with a transition rate per unit of volume
\bea 
\Gamma_{\rm tunnel} \sim  m_\phi^4e^{-S_4}~,
\eea 
where $S_4$ is the Euclidean action of the bounce profile controlling the tunneling. Vacuum stability until today is ensured only if $\Gamma_{\rm tun} \ll 1/V_0t_0 \sim H_0^4 \Rightarrow S_4\gtrsim 400$.

The first possibility is that the tunnelling occurs from the local trapping minimum $\phi_{\rm min} \sim \phi_c + f$ towards the true minimum at $\phi=0$. The Euclidean action is estimated using the triangular approximation~\cite{DUNCAN1992109,Amariti:2020ntv}
\bea 
S_4^{\rm trg} = \frac{32\pi^2}{3}\frac{1+c}{(\sqrt{1+c}-1)^4} \left(\frac{\Delta \phi_+^4}{\Delta V_+}\right)~,
\eea 
where \bea 
&\Delta V_+ \sim M^4~, \quad \Delta \phi_+ \sim f~,\quad 
\notag \\
&\Delta V_- \sim g^4\frac{\phi_c^4}{\lambda}~, \quad  \Delta \phi_- \sim \phi_c~, 
\notag \\
 &c \equiv \frac{\Delta V_-}{\Delta V_+}\frac{ \Delta \phi_+}{ \Delta \phi_-} \approx \frac{g^4}{\lambda} \left(\frac{\phi_c}{M}\right)^4 \frac{ f}{\phi_c} =  \frac{g \Lambda^3  f}{\lambda M^4}~.
\eea 
We denote as $\Delta \phi_{+(-)}$ the distance in field space between the local maximum and the false (true) minimum and $\Delta V_{+(-)}$ the difference of the values of the potential between those points.

In the limit $c\gg 1$, the Euclidean action becomes
\bea  \label{eq:condi}
S_4^{\rm trg} \approx \frac{32\pi^2}{3}  \frac{\lambda f^3}{g \Lambda^3}~.
\eea 
The condition $S_4\gtrsim 400$ leads to
\bea 
\frac{\lambda f^3}{g \Lambda^3} \gtrsim 5~.
\label{eq:triang_bound}
\eea 
Then we need to consider the case where the tunnelling occurs regionally towards another minimum in the vicinity. Due to the fact that the values of the potential between subsequent barriers are almost equal, here it is more suitable to use the rectangular approximation\cite{DUNCAN1992109,Amariti:2020ntv}, which is given by 
\bea \label{eq:rectang}
S_4^{\rm rcg} = 2\pi^2 \frac{\Delta \phi^4}{\left(\Delta V_+^{1/3}-\Delta V_-^{1/3}\right)^3}~,
\eea 
where 
\bea \label{eq:rectang_terms}
\Delta \phi \sim f~, \quad  \Delta V_- &\sim M^4- \frac{g^2 \Lambda^2 (f)^2}{\lambda} - \frac{g \Lambda^3 f}{8\pi^2}~,\quad 
\notag \\
\Delta V_+ &\sim M^4+ \frac{g^2 \Lambda^2 f^2}{\lambda} + \frac{g \Lambda^3  f}{8\pi^2}~.
\eea 
In the regime of Eq. (4) (main text), the first term in Eq. \eqref{eq:rectang_terms} dominates and neglecting the second term, we obtain 
\bea 
S_4^{\rm rcg} \approx 27648 \pi^8  \frac{f M^8}{g^3 \Lambda^9}~,
\label{eq:applica}
\eea 
which implies that $S_4^{\rm rcg}\gg S_4^{\rm trg}$ and thus this transition is suppressed in comparison to the one towards the true minimum.

\hspace{1cm}

{\small
\bibliographystyle{apsrev4-2}

\begin{thebibliography}{0}%
\makeatletter
\providecommand \@ifxundefined [1]{%
 \@ifx{#1\undefined}
}%
\providecommand \@ifnum [1]{%
 \ifnum #1\expandafter \@firstoftwo
 \else \expandafter \@secondoftwo
 \fi
}%
\providecommand \@ifx [1]{%
 \ifx #1\expandafter \@firstoftwo
 \else \expandafter \@secondoftwo
 \fi
}%
\providecommand \natexlab [1]{#1}%
\providecommand \enquote  [1]{``#1''}%
\providecommand \bibnamefont  [1]{#1}%
\providecommand \bibfnamefont [1]{#1}%
\providecommand \citenamefont [1]{#1}%
\providecommand \href@noop [0]{\@secondoftwo}%
\providecommand \href [0]{\begingroup \@sanitize@url \@href}%
\providecommand \@href[1]{\@@startlink{#1}\@@href}%
\providecommand \@@href[1]{\endgroup#1\@@endlink}%
\providecommand \@sanitize@url [0]{\catcode `\\12\catcode `\$12\catcode
  `\&12\catcode `\#12\catcode `\^12\catcode `\_12\catcode `\%12\relax}%
\providecommand \@@startlink[1]{}%
\providecommand \@@endlink[0]{}%
\providecommand \url  [0]{\begingroup\@sanitize@url \@url }%
\providecommand \@url [1]{\endgroup\@href {#1}{\urlprefix }}%
\providecommand \urlprefix  [0]{URL }%
\providecommand \Eprint [0]{\href }%
\providecommand \doibase [0]{https://doi.org/}%
\providecommand \selectlanguage [0]{\@gobble}%
\providecommand \bibinfo  [0]{\@secondoftwo}%
\providecommand \bibfield  [0]{\@secondoftwo}%
\providecommand \translation [1]{[#1]}%
\providecommand \BibitemOpen [0]{}%
\providecommand \bibitemStop [0]{}%
\providecommand \bibitemNoStop [0]{.\EOS\space}%
\providecommand \EOS [0]{\spacefactor3000\relax}%
\providecommand \BibitemShut  [1]{\csname bibitem#1\endcsname}%
\let\auto@bib@innerbib\@empty
\end{thebibliography}%


\begin{thebibliography}{72}%
\makeatletter
\providecommand \@ifxundefined [1]{%
 \@ifx{#1\undefined}
}%
\providecommand \@ifnum [1]{%
 \ifnum #1\expandafter \@firstoftwo
 \else \expandafter \@secondoftwo
 \fi
}%
\providecommand \@ifx [1]{%
 \ifx #1\expandafter \@firstoftwo
 \else \expandafter \@secondoftwo
 \fi
}%
\providecommand \natexlab [1]{#1}%
\providecommand \enquote  [1]{``#1''}%
\providecommand \bibnamefont  [1]{#1}%
\providecommand \bibfnamefont [1]{#1}%
\providecommand \citenamefont [1]{#1}%
\providecommand \href@noop [0]{\@secondoftwo}%
\providecommand \href [0]{\begingroup \@sanitize@url \@href}%
\providecommand \@href[1]{\@@startlink{#1}\@@href}%
\providecommand \@@href[1]{\endgroup#1\@@endlink}%
\providecommand \@sanitize@url [0]{\catcode `\\12\catcode `\$12\catcode
  `\&12\catcode `\#12\catcode `\^12\catcode `\_12\catcode `\%12\relax}%
\providecommand \@@startlink[1]{}%
\providecommand \@@endlink[0]{}%
\providecommand \url  [0]{\begingroup\@sanitize@url \@url }%
\providecommand \@url [1]{\endgroup\@href {#1}{\urlprefix }}%
\providecommand \urlprefix  [0]{URL }%
\providecommand \Eprint[0]{\href }%
\providecommand \doibase [0]{https://doi.org/}%
\providecommand \selectlanguage [0]{\@gobble}%
\providecommand \bibinfo  [0]{\@secondoftwo}%
\providecommand \bibfield  [0]{\@secondoftwo}%
\providecommand \translation [1]{[#1]}%
\providecommand \BibitemOpen [0]{}%
\providecommand \bibitemStop [0]{}%
\providecommand \bibitemNoStop [0]{.\EOS\space}%
\providecommand \EOS [0]{\spacefactor3000\relax}%
\providecommand \BibitemShut  [1]{\csname bibitem#1\endcsname}%
\let\auto@bib@innerbib\@empty

\bibitem{ATLAS:2012yve}
{\bf ATLAS} Collaboration, G.~Aad et~al. {\em Phys. Lett. B} {\bf 716} (2012)
  1--29, [\href{http://arxiv.org/abs/1207.7214}{{\tt arXiv:1207.7214}}].

\bibitem{Arvanitaki:2013yja}
A.~Arvanitaki, M.~Baryakhtar, X.~Huang, K.~van Tilburg, and G.~Villadoro {\em
  JHEP} {\bf 03} (2014) 022, [\href{http://arxiv.org/abs/1309.3568}{{\tt
  arXiv:1309.3568}}].

\bibitem{Sanz:2017tco}
V.~Sanz and J.~Setford {\em Adv. High Energy Phys.} {\bf 2018} (2018) 7168480,
  [\href{http://arxiv.org/abs/1703.10190}{{\tt arXiv:1703.10190}}].

\bibitem{Baer:2020kwz}
H.~Baer, V.~Barger, S.~Salam, D.~Sengupta, and K.~Sinha {\em Eur. Phys. J. ST}
  {\bf 229} (2020), no.~21 3085--3141,
  [\href{http://arxiv.org/abs/2002.03013}{{\tt arXiv:2002.03013}}].

\bibitem{Graham:2015cka}
P.~W. Graham, D.~E. Kaplan, and S.~Rajendran {\em Phys. Rev. Lett.} {\bf 115}
  (2015), no.~22 221801, [\href{http://arxiv.org/abs/1504.07551}{{\tt
  arXiv:1504.07551}}].

\bibitem{Hardy:2015laa}
E.~Hardy {\em JHEP} {\bf 11} (2015) 077,
  [\href{http://arxiv.org/abs/1507.07525}{{\tt arXiv:1507.07525}}].

\bibitem{Fowlie:2016jlx}
A.~Fowlie, C.~Balazs, G.~White, L.~Marzola, and M.~Raidal {\em JHEP} {\bf 08}
  (2016) 100, [\href{http://arxiv.org/abs/1602.03889}{{\tt arXiv:1602.03889}}].

\bibitem{Espinosa:2015eda}
J.~R. Espinosa, C.~Grojean, G.~Panico, A.~Pomarol, O.~Pujol\`as, and G.~Servant
  {\em Phys. Rev. Lett.} {\bf 115} (2015), no.~25 251803,
  [\href{http://arxiv.org/abs/1506.09217}{{\tt arXiv:1506.09217}}].

\bibitem{Gupta:2015uea}
R.~S. Gupta, Z.~Komargodski, G.~Perez, and L.~Ubaldi {\em JHEP} {\bf 02} (2016)
  166, [\href{http://arxiv.org/abs/1509.00047}{{\tt arXiv:1509.00047}}].

\bibitem{Choi:2015fiu}
K.~Choi and S.~H. Im {\em JHEP} {\bf 01} (2016) 149,
  [\href{http://arxiv.org/abs/1511.00132}{{\tt arXiv:1511.00132}}].

\bibitem{Kaplan:2015fuy}
D.~E. Kaplan and R.~Rattazzi {\em Phys. Rev. D} {\bf 93} (2016), no.~8 085007,
  [\href{http://arxiv.org/abs/1511.01827}{{\tt arXiv:1511.01827}}].

\bibitem{Ibanez:2015fcv}
L.~E. Ibanez, M.~Montero, A.~Uranga, and I.~Valenzuela {\em JHEP} {\bf 04}
  (2016) 020, [\href{http://arxiv.org/abs/1512.00025}{{\tt arXiv:1512.00025}}].

\bibitem{Hebecker:2015zss}
A.~Hebecker, F.~Rompineve, and A.~Westphal {\em JHEP} {\bf 04} (2016) 157,
  [\href{http://arxiv.org/abs/1512.03768}{{\tt arXiv:1512.03768}}].

\bibitem{Fonseca:2016eoo}
N.~Fonseca, L.~de~Lima, C.~S. Machado, and R.~D. Matheus {\em Phys. Rev. D}
  {\bf 94} (2016), no.~1 015010, [\href{http://arxiv.org/abs/1601.07183}{{\tt
  arXiv:1601.07183}}].

\bibitem{Evans:2016htp}
J.~L. Evans, T.~Gherghetta, N.~Nagata, and Z.~Thomas {\em JHEP} {\bf 09} (2016)
  150, [\href{http://arxiv.org/abs/1602.04812}{{\tt arXiv:1602.04812}}].

\bibitem{Hook:2016mqo}
A.~Hook and G.~Marques-Tavares {\em JHEP} {\bf 12} (2016) 101,
  [\href{http://arxiv.org/abs/1607.01786}{{\tt arXiv:1607.01786}}].

\bibitem{Choi:2016kke}
K.~Choi, H.~Kim, and T.~Sekiguchi {\em Phys. Rev. D} {\bf 95} (2017), no.~7
  075008, [\href{http://arxiv.org/abs/1611.08569}{{\tt arXiv:1611.08569}}].

\bibitem{Tangarife:2017rgl}
W.~Tangarife, K.~Tobioka, L.~Ubaldi, and T.~Volansky {\em JHEP} {\bf 02} (2018)
  084, [\href{http://arxiv.org/abs/1706.03072}{{\tt arXiv:1706.03072}}].

\bibitem{Wang:2018ddr}
S.-J. Wang {\em Phys. Rev. D} {\bf 99} (2019), no.~9 095026,
  [\href{http://arxiv.org/abs/1811.06520}{{\tt arXiv:1811.06520}}].

\bibitem{Fonseca:2018xzp}
N.~Fonseca, E.~Morgante, and G.~Servant {\em JHEP} {\bf 10} (2018) 020,
  [\href{http://arxiv.org/abs/1805.04543}{{\tt arXiv:1805.04543}}].

\bibitem{Ibe:2019udh}
M.~Ibe, Y.~Shoji, and M.~Suzuki {\em JHEP} {\bf 11} (2019) 140,
  [\href{http://arxiv.org/abs/1904.02545}{{\tt arXiv:1904.02545}}].

\bibitem{Fonseca:2019lmc}
N.~Fonseca, E.~Morgante, R.~Sato, and G.~Servant {\em JHEP} {\bf 05} (2020)
  080, [\href{http://arxiv.org/abs/1911.08473}{{\tt arXiv:1911.08473}}].
  [Erratum: JHEP 01, 012 (2021)].

\bibitem{Domcke:2021yuz}
V.~Domcke, K.~Schmitz, and T.~You {\em JHEP} {\bf 07} (2022) 126,
  [\href{http://arxiv.org/abs/2108.11295}{{\tt arXiv:2108.11295}}].

\bibitem{Klangburam:2022igc}
T.~Klangburam, A.~Waeming, P.~Tantirangsri, D.~Samart, and C.~Pongkitivanichkul
  {\em JHEP} {\bf 06} (2022) 159, [\href{http://arxiv.org/abs/2202.08857}{{\tt
  arXiv:2202.08857}}].

\bibitem{Chatrchyan:2022pcb}
A.~Chatrchyan and G.~Servant \href{http://arxiv.org/abs/2210.01148}{{\tt
  arXiv:2210.01148}}.

\bibitem{Geller:2018xvz}
M.~Geller, Y.~Hochberg, and E.~Kuflik {\em Phys. Rev. Lett.} {\bf 122} (2019),
  no.~19 191802, [\href{http://arxiv.org/abs/1809.07338}{{\tt
  arXiv:1809.07338}}].

\bibitem{Cheung:2018xnu}
C.~Cheung and P.~Saraswat \href{http://arxiv.org/abs/1811.12390}{{\tt
  arXiv:1811.12390}}.

\bibitem{TitoDAgnolo:2021nhd}
R.~Tito~D'Agnolo and D.~Teresi {\em Phys. Rev. Lett.} {\bf 128} (2022), no.~2
  021803, [\href{http://arxiv.org/abs/2106.04591}{{\tt arXiv:2106.04591}}].

\bibitem{Kofman:2004yc}
L.~Kofman, A.~D. Linde, X.~Liu, A.~Maloney, L.~McAllister, and E.~Silverstein
  {\em JHEP} {\bf 05} (2004) 030,
  [\href{http://arxiv.org/abs/hep-th/0403001}{{\tt hep-th/0403001}}].

\bibitem{PhysRevD.80.063533}
D.~Green, B.~Horn, L.~Senatore, and E.~Silverstein {\em Phys. Rev. D} {\bf 80}
  (Sep, 2009) 063533.

\bibitem{Pearce:2016qtn}
L.~Pearce, M.~Peloso, and L.~Sorbo {\em JCAP} {\bf 11} (2016) 058,
  [\href{http://arxiv.org/abs/1603.08021}{{\tt arXiv:1603.08021}}].

\bibitem{Dimopoulos:2019ogl}
K.~Dimopoulos, M.~Kar\v{c}iauskas, and C.~Owen {\em Phys. Rev. D} {\bf 100}
  (2019), no.~8 083530, [\href{http://arxiv.org/abs/1907.04676}{{\tt
  arXiv:1907.04676}}].

\bibitem{Karciauskas:2021fdu}
M.~Kar\v{c}iauskas, S.~Rusak, and A.~Saez {\em Phys. Rev. D} {\bf 105} (2022),
  no.~4 043535, [\href{http://arxiv.org/abs/2112.11536}{{\tt
  arXiv:2112.11536}}].

\bibitem{Antipin:2015jia}
O.~Antipin and M.~Redi {\em JHEP} {\bf 12} (2015) 031,
  [\href{http://arxiv.org/abs/1508.01112}{{\tt arXiv:1508.01112}}].

\bibitem{Kofman:1997yn}
L.~Kofman, A.~D. Linde, and A.~A. Starobinsky {\em Phys. Rev. D} {\bf 56}
  (1997) 3258--3295, [\href{http://arxiv.org/abs/hep-ph/9704452}{{\tt
  hep-ph/9704452}}].

\bibitem{Dufaux:2006ee}
J.~F. Dufaux, G.~N. Felder, L.~Kofman, M.~Peloso, and D.~Podolsky {\em JCAP}
  {\bf 07} (2006) 006, [\href{http://arxiv.org/abs/hep-ph/0602144}{{\tt
  hep-ph/0602144}}].

\bibitem{Dufaux:2008dn}
J.-F. Dufaux, G.~Felder, L.~Kofman, and O.~Navros {\em JCAP} {\bf 03} (2009)
  001, [\href{http://arxiv.org/abs/0812.2917}{{\tt arXiv:0812.2917}}].

\bibitem{Abolhasani:2009nb}
A.~A. Abolhasani, H.~Firouzjahi, and M.~M. Sheikh-Jabbari {\em Phys. Rev. D}
  {\bf 81} (2010) 043524, [\href{http://arxiv.org/abs/0912.1021}{{\tt
  arXiv:0912.1021}}].

\bibitem{Fedderke:2014ura}
M.~A. Fedderke, E.~W. Kolb, and M.~Wyman {\em Phys. Rev. D} {\bf 91} (2015),
  no.~6 063505, [\href{http://arxiv.org/abs/1409.1584}{{\tt arXiv:1409.1584}}].

\bibitem{ATLAS:2016neq}
{\bf ATLAS, CMS} Collaboration, G.~Aad et~al. {\em JHEP} {\bf 08} (2016) 045,
  [\href{http://arxiv.org/abs/1606.02266}{{\tt arXiv:1606.02266}}].

\bibitem{Ilnicka:2018def}
A.~Ilnicka, T.~Robens, and T.~Stefaniak {\em Mod. Phys. Lett. A} {\bf 33}
  (2018), no.~10n11 1830007, [\href{http://arxiv.org/abs/1803.03594}{{\tt
  arXiv:1803.03594}}].
	
\bibitem{Adhikari:2020vqo}
S.~Adhikari, I.~M.~Lewis and M.~Sullivan, {\em Phys. Rev. D} {\bf 103} (2021),
 no.~7 075027, [\href{http://arxiv.org/abs/2003.10449}{{\tt arXiv:2003.10449}}].

\bibitem{EuropeanStrategyforParticlePhysicsPreparatoryGroup:2019qin}
R.~K. Ellis et~al. \href{http://arxiv.org/abs/1910.11775}{{\tt
  arXiv:1910.11775}}.

\bibitem{ATLAS:2017otj}
{\bf ATLAS} Collaboration, M.~Aaboud et~al. {\em JHEP} {\bf 03} (2018) 009,
  [\href{http://arxiv.org/abs/1708.09638}{{\tt arXiv:1708.09638}}].

\bibitem{CMS:2018amk}
{\bf CMS} Collaboration, A.~M. Sirunyan et~al. {\em JHEP} {\bf 06} (2018) 127,
  [\href{http://arxiv.org/abs/1804.01939}{{\tt arXiv:1804.01939}}]. [Erratum:
  JHEP 03, 128 (2019)].

\bibitem{Buttazzo:2018qqp}
D.~Buttazzo, D.~Redigolo, F.~Sala, and A.~Tesi {\em JHEP} {\bf 11} (2018) 144,
  [\href{http://arxiv.org/abs/1807.04743}{{\tt arXiv:1807.04743}}].

\bibitem{MuonCollider:2022xlm}
{\bf Muon Collider} Collaboration, J.~de~Blas et~al.
  \href{http://arxiv.org/abs/2203.07261}{{\tt arXiv:2203.07261}}.

\bibitem{PhysRevD.28.1243}
M.~S. Turner {\em Phys. Rev. D} {\bf 28} (Sep, 1983) 1243--1247.

\bibitem{DUNCAN1992109}
M.~Duncan and L.~G. Jensen {\em Physics Letters B} {\bf 291} (1992), no.~1
  109--114.

\bibitem{Amariti:2020ntv}
A.~Amariti \href{http://arxiv.org/abs/2009.14102}{{\tt arXiv:2009.14102}}.
\end{thebibliography}
}
\end{document}